\documentclass[journal,twoside,web]{ieeecolor}
\usepackage{tmi}
\usepackage{cite}
\usepackage[table]{xcolor}
\usepackage{amsmath,amssymb,amsfonts}
\usepackage{graphicx}
\usepackage{textcomp}

\usepackage{framed,multirow}

\usepackage{algorithm, algpseudocode}

\usepackage{moreverb}
\usepackage[utf8]{inputenc}
\usepackage{notation}
\usepackage{color}

\usepackage{booktabs}

\begin{document}

\title{ Unsupervised stratification of patients with myocardial infarction based on imaging and in-silico biomarkers}
 
\author{Dolors Serra, Pau Romero, Paula Franco, Ignacio Bernat, Miguel Lozano, Ignacio Garcia-Fernandez, David Soto, Antonio Berruezo, Oscar Camara and Rafael Sebastian
\thanks{This work has been submitted to the IEEE for possible publication. Copyright may be transferred without notice, after which this version may no longer be accessible.
Corresponding author: Miguel Lozano. This research was funded by Generalitat Valenciana Grant AICO/2021/318 (Consolidables 2021) and Grant PID2020-114291RB-I00 funded by MCIN/ 10.13039/501100011033 and, by "ERDF A way of making Europe"; 
D.S. was funded by the Generalitat Valenciana and the European Social Fund (FSE) through the Recruitment of Predoctoral Research Staff ACIF/2021/394 included in the FSE Operational Program 2021-2025 of the Valencian Community (Spain);.}
\thanks{D. Serra, P. Romero, M. Lozano, I. Garcia-Fernandez and R. Sebastian are with Computational Multiscale Simulation Lab (CoMMLab), Departament d'Informatica, Universitat de Valencia, 46100 Burjassot, Spain (email: rafael.sebastian@uv.es)}
\thanks{D. Soto and A. Berruezo Cardiology Department, Heart Institute, Teknon Medical Center, Barcelona, Spain}
}

\maketitle

\begin{abstract}
This study presents a novel methodology for stratifying post-myocardial infarction patients at risk of ventricular arrhythmias using patient-specific 3D cardiac models derived from late gadolinium enhancement cardiovascular magnetic resonance (LGE-CMR) images. The method integrates imaging and computational simulation with a simplified cellular automaton model, Arrhythmic3D, enabling rapid and accurate VA risk assessment in clinical timeframes. Applied to 51 patients, the model generated thousands of personalized simulations to evaluate arrhythmia inducibility and predict VA risk. Key findings include the identification of slow conduction channels (SCCs) within scar tissue as critical to reentrant arrhythmias and the localization of high-risk zones for potential intervention. The Arrhythmic Risk Score (ARRISK), developed from simulation results, demonstrated strong concordance with clinical outcomes and outperformed traditional imaging-based risk stratification. The methodology is fully automated, requiring minimal user intervention, and offers a promising tool for improving precision medicine in cardiac care by enhancing patient-specific arrhythmia risk assessment and guiding treatment strategies.
\end{abstract}



\section{Introduction}
\IEEEPARstart{P}{redicting} ventricular arrhythmias in post-myocardial infarction patients, especially in the early phase, remains a significant clinical challenge due to the elevated risk of sudden death. Early implantable cardioverter-defibrillator placement does not significantly improve mortality~\cite{Pouleur:2010aa}.
Postinfarct ventricular tachycardia typically arises as a reentrant arrhythmia sustained by circuits formed by surviving myocardial strands in scar tissue, where slow conduction channels (SCC) exhibit reduced gap junction density and impaired excitability~\cite{Donahue:2024aa}. LGE-CMR can identify SCCs to guide catheter ablation, though it remains challenging due to the difficulty in locating optimal targets and the potential for extensive lesions~\cite{stevenson1998radiofrequency,Cronin:2020aa}.

Precision medicine in clinical prognosis aims to provide personalized therapies for each patient. The concept of a digital twin enables the integration of patient-specific clinical data~\cite{Trayanova:2024aa}. In cardiology, combining patient-specific simulations with clinical metrics can improve the prediction of ventricular arrhythmia risks and enhance the identification of problematic areas~\cite{Corral-Acero:2020aa}.

Advanced biophysical models simulate cellular electrical dynamics, pathological changes, and drug responses~\cite{ten2007organization,Niederer:2011wg}. These models are used for simulating VTs due to their ability to model ischemic tissue's heterogeneous properties~\cite{Chen:2016aa, Deng:2019aa, Lopez-Perez:2019aa, Prakosa:2018aa, Zhang:2023aa}. However, they require high spatial and temporal discretization, leading to lengthy computation times. Despite advances in multi-GPU computations, the complexity and hardware demands limit clinical application. Simplified computational models have emerged to address this issue, offering accurate results suitable for clinical use~\cite{Bhagirath:2023aa, Loewe:2018uz, Neic:2017um}.

To integrate imaging with computer simulation technology in clinical practice, we propose a methodology for estimating ventricular arrhythmia risk using image-based personalized 3D models that can be executed in clinical timeframes. This approach utilizes Arrhythmic3D, a physiological cellular automaton derived from thousands of pre-computed biophysical simulations, coupled with an Eikonal model~\cite{serra:2022au}. It accurately reproduces cellular and tissue heterogeneity with low spatial and temporal discretization, allowing for rapid computation on desktop computers.

This study demonstrates the potential of simplified electrophysiological simulations for arrhythmia risk assessment, applied to a cohort of 51 myocardial infarction patients. The use of Arrhythmic3D enabled thousands of patient-specific simulations, covering various risk scenarios and individual variabilities. The method is fully automated post-segmentation, requiring no manual intervention, and completes in minutes, highlighting its efficiency and clinical applicability.
The research details the methodology, presents simulation results, and compares the predicted arrhythmia risk with clinical follow-up data.

\section{Material and methods}

\subsection{Patient dataset}

Patients with chronic myocardial infarction were recruited at the Teknon Clinical Center (Barcelona, Spain), where they underwent late gadolinium enhancement cardiovascular magnetic resonance (LGE-CMR) scans for detailed heart imaging. The images were acquired using a 1.5-Tesla scanner (MAGNETOM TrioTM, Siemens Healthcare, Erlangen, Germany) with a resolution of 1 mm between planes and an isotropic in-plane resolution of 0.9 mm, producing 120 slices per scan. The LGE-CMR scans utilized the 3D-DIXON imaging technique. This high resolution is crucial for accurately reconstructing pathological tissue in patient-specific computational cardiac models~\cite{Mendoca:2021aa}.

For each patient, the clinical data includes whether the patient experienced VT, whether VT inducibility was attempted in the clinic, and if so, whether the attempt was successful. Additionally, a risk classification is provided. VT episodes are categorized as HIGH risk when identified as sustained monomorphic VT (SMVT, characterized by regular QRS morphology) or Polymorphic VT (with varying QRS complex morphology) due to their potential severity. They are classified as LOW risk when identified as non-sustained VT (NSVT), and as ZERO risk when VT was neither experienced nor induced. The complete cohort data is available in Table~\ref{tb:datostotal}. 



\subsection{Digital Twin Construction}
\begin{figure}[!h]
    \begin{center}
    \includegraphics[width=\linewidth]{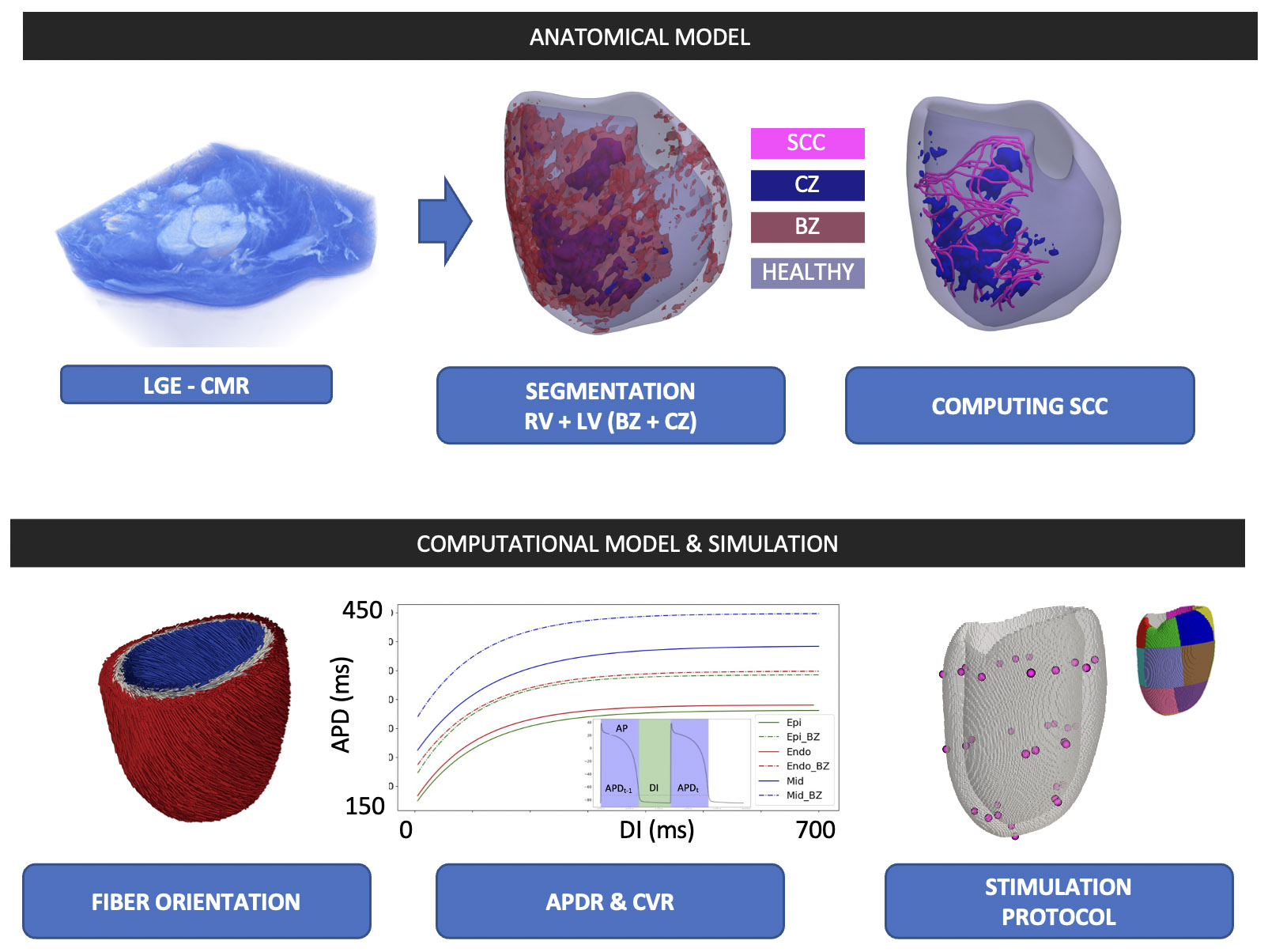} 
    \end{center}
    \caption{LGE-CMR-based Digital Twin construction workflow and data preprocessing for simulations }
    \label{fig:pipe}
\end{figure}

The anatomical model for each patient was constructed through the segmentation of LGE-CMR image scans. The ADAS3D software (Galgo Medical, Barcelona) was used to segment the endocardial and epicardial surfaces, defining the closed heart boundaries. The scar region was segmented using the maximum intensity pixel method, as implemented in ADAS3D, with a default 40-60 pixel-signal-intensity (PSI) threshold under expert supervision. This process allowed for differentiation between the border zone (BZ), scar core zone (CZ), and healthy tissue within the scar region. The entire segmentation process took approximately 10 minutes per patient. From these LGE-CMR scans, volumetric models were generated, and clinical biomarkers were calculated (summarized data available in Fig.~\ref{fig:pipe}).
Given that SCCs are critical paths involved in the maintenance of VT, we processed the scar region to extract the center lines of the BZ areas surrounded by CZ tissue, aiming to characterize all potential reentry paths.

Using the segmented surfaces, we constructed a digital twin for each patient. The voxelization of the models was directly derived from the LGE-CMR stacks, which provided an isotropic resolution of approximately 1 mm. Subsequently, the nodes and elements of the models were labeled according to the cellular populations located in the endocardial, midmyocardial, and epicardial regions, based on their position across the myocardial wall~\cite{Lopez-Perez:2015aa}. Labels for healthy, BZ, or CZ tissue were applied to adjust the conduction velocity restitution (CVR) and action potential duration restitution (APDR) curves accordingly.

\begin{figure}[!h]
    \begin{center}
    \includegraphics[width=0.9\linewidth]{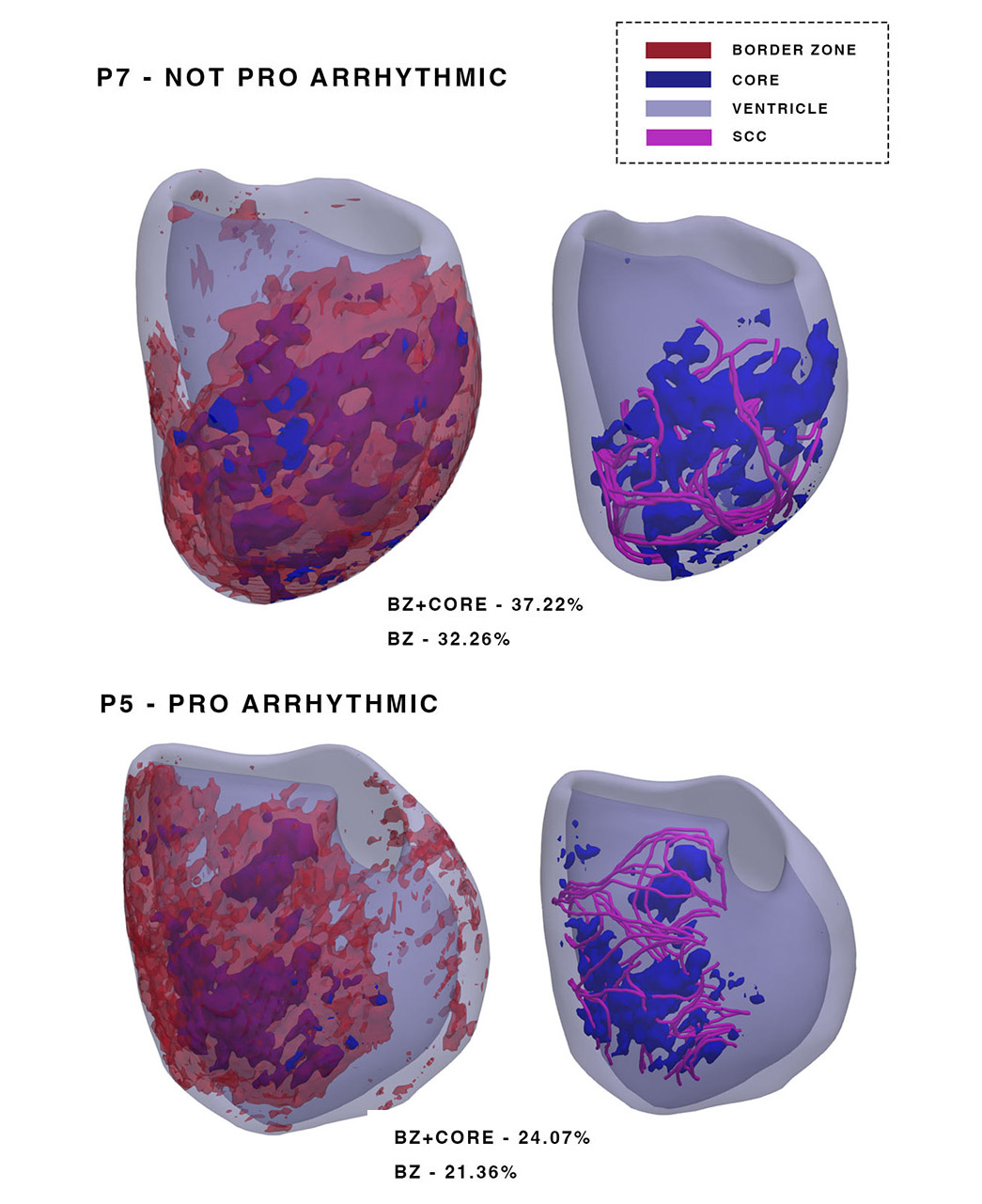} 
    \end{center}
    \caption{Segmented models for patients P7 and P5 featuring personalized anatomy, including scar, border Zone, core zone and the extraction of slow conduction channels that extend across the core zone.}
    \label{fig:segmented-models}
\end{figure}

All clinical cases exhibited complex scar and SCC structures (represented by magenta tubes in Fig.~\ref{fig:segmented-models}), regardless of whether the patients had experienced clinical arrhythmic events. Fiber orientation was determined using a rule-based model, specifically the Streeter model~\cite{streeter1969fiber}, and was incorporated as a property of the model's elements.
It is important to note that the entire post-segmentation process for creating the digital twins was fully automated, requiring no manual user intervention and taking approximately 2 minutes per case.

\onecolumn
\renewcommand{\arraystretch}{1.15}
\begin{table}[!h]
    \centering
    \begin{tabular}{|p{0.035\linewidth}|p{0.05\linewidth}|c|c|c|p{0.04\linewidth}|c|c|c|p{0.03\linewidth}|p{0.05\linewidth}|p{0.03\linewidth}|p{0.03\linewidth}|p{0.06\linewidth}|p{0.05\linewidth}|}
    \hline
        \textbf{PAT.} & \textbf{LV MASS (g)} & \textbf{BZ (g)} & \textbf{CZ (g)} & \textbf{SCC(g)} & \textbf{LVEF (\%)} & \textbf{AGE} & \textbf{SEX} & \textbf{VT } & \textbf{Sim Time (h)} & \textbf{Effective sims} & \textbf{\#VT} & \textbf{\#VT base-line} & \textbf{ARRISK} & \textbf{Clinical risk} \\ \hline
        \textbf{P1} & 82.18 & 4.45 & 0.64 & 0 & 66 & 71 & M & N-X-N & 15 & 1998 & 0 & 0 & ZERO & ZERO \\  \hline
         \rowcolor{green!25}
        \textbf{P2} & 109.10 & 13.36 & 3.48 & 8.05 & 36 & 53 & M & P-P-P & 20 & 1934 & 22 & 16 & \cellcolor{red!15}HIGH & HIGH \\  \hline
        \textbf{P3} & 100.88 & 4.65 & 0.51 & 0 & 61 & 63 & M & N-X-N & 22 & 1884 & 0 & 0 & ZERO & ZERO \\  \hline
        \textbf{P4} & 67.81 & 3.50 & 2.48 & 2.37 & 63 & 74 & F & N-X-P & 10 & 1506 & 1 & 1 & LOW & ZERO \\  \hline
         \rowcolor{green!25}
        \textbf{P5} & 197.65 & 42.21 & 5.36 & 21.09 & 28 & 84 & M & N-P-P & 27 & 1830 & 23 & 10 & \cellcolor{red!15}HIGH & HIGH \\  \hline
        \textbf{P6} & 73.46 & 2.74 & 0.56 & 0.55 & 69 & 48 & F & N-X-N & 12 & 1812 & 0 & 0 & ZERO & ZERO \\  \hline
        \textbf{P7} & 79.80 & 25.74 & 3.96 & 14.94 & 47 & 78 & F & N-X-P & 12 & 1656 & 10 & 9 & LOW & ZERO \\  \hline
        \textbf{P8} & 90.65 & 1.98 & 0.56 & 0 & 58 & 56 & M & N-X-N & 14 & 1787 & 0 & 0 & ZERO & ZERO \\  \hline
        \textbf{P9} & 111.51 & 10.35 & 3.91 & 1.94 & 63 & 78 & M & N-X-P & 19 & 1966 & 5 & 3 & LOW & ZERO \\  \hline
        \textbf{P10} & 120.34 & 5.56 & 1.50 & 0.85 & 63 & 68 & M & N-X-P & 22 & 1764 & 1 & 1 & LOW & ZERO \\  \hline
        \textbf{P11} & 109.81 & 9.22 & 2.48 & 0.64 & 30 & 54 & M & N-X-N & 15 & 1818 & 0 & 0 & ZERO & ZERO \\  \hline
         \rowcolor{green!25}
        \textbf{P12} & 115.85 & 24.25 & 12.58 & 9.06 & 47 & 70 & M & P-X-P & 18 & 1830 & 7 & 3 & LOW & LOW \\  \hline
        \textbf{P13} & 81.80 & 2.48 & 0.75 & 1.28 & 63 & 58 & M & N-X-N & 13 & 1644 & 0 & 0 & ZERO & ZERO \\  \hline
        \textbf{P14} & 179.37 & 53.20 & 35.41 & 35.19 & 40 & 39 & M & N-X-P & 20 & 1782 & 27 & 14 & \cellcolor{red!15}HIGH & ZERO \\  \hline
        \textbf{P15} & 100.79 & 8.44 & 1.15 & 0 & 59 & 69 & M & N-X-N & 19 & 1860 & 0 & 0 & ZERO & ZERO \\  \hline
        \textbf{P16} & 195.69 & 27.72 & 12.72 & 2.33 & 41 & 66 & M & N-X-P & 26 & 1938 & 13 & 10 & LOW & ZERO \\  \hline
        \textbf{P18} & 146.47 & 7.11 & 4.65 & 2.35 & 46 & 63 & M & N-X-P & 26 & 1926 & 2 & 1 & LOW & ZERO \\  \hline
        \textbf{P19} & 89.10 & 6.59 & 4.53 & 0.63 & 48 & 49 & M & N-X-N & 13 & 1884 & 0 & 0 & ZERO & ZERO \\  \hline
        \textbf{P20} & 151.43 & 3.92 & 1.87 & 0 & 44 & 41 & M & N-X-N & 24 & 1998 & 0 & 0 & ZERO & ZERO \\  \hline
        \textbf{P21} & 96.76 & 8.04 & 2.03 & 2.02 & 60 & 56 & M & N-X-P & 17 & 1990 & 2 & 2 & LOW & ZERO \\  \hline
         \rowcolor{green!25}
        \textbf{P23} & 105.94 & 13.76 & 4.82 & 3.59 & 57 & 63 & M & P-N-P & 20 & 2020 & 3 & 3 & LOW & LOW \\  \hline
        \textbf{P24} & 110.26 & 14.78 & 3.91 & 10.21 & 47 & 64 & M & N-X-P & 13 & 1788 & 1 & 1 & LOW & ZERO \\  \hline
        \textbf{P26} & 137.54 & 13.22 & 6.21 & 2.81 & 57 & 61 & M & N-X-P & 35 & 2022 & 26 & 16 & \cellcolor{red!15}HIGH & ZERO \\  \hline
        \textbf{P28} & 60.24 & 1.12 & 0.13 & 0 & 56 & 47 & F & N-X-P & 19 & 1860 & 5 & 0 & LOW & ZERO \\  \hline
        \textbf{P29} & 108.39 & 20.34 & 3.67 & 14.35 & 55 & 80 & F & N-X-P & 19 & 1914 & 2 & 2 & LOW & ZERO \\  \hline
        \textbf{P30} & 139.80 & 8.10 & 9.02 & 0.24 & 43 & 75 & F & N-X-N & 17 & 1956 & 0 & 0 & ZERO & ZERO \\  \hline
        \textbf{P32} & 123.38 & 18.55 & 4.11 & 7.97 & 49 & 73 & F & N-X-P & 18 & 1842 & 4 & 2 & LOW & ZERO \\  \hline
        \textbf{P33} & 105.41 & 17.02 & 1.59 & 5.72 & 57 & 49 & F & N-X-P & 17 & 1962 & 1 & 1 & LOW & ZERO \\  \hline
        \textbf{P34} & 87.28 & 11.80 & 1.75 & 1.28 & 60 & 70 & M & N-X-N & 11 & 1560 & 0 & 0 & ZERO & ZERO \\  \hline
        \textbf{P35} & 115.95 & 18.06 & 5.99 & 8.39 & 65 & 73 & M & N-X-P & 14 & 1956 & 1 & 1 & LOW & ZERO \\  \hline
        \textbf{P36} & 101.41 & 22.24 & 4.74 & 15.14 & 32 & 61 & M & N-X-P & 15 & 2010 & 6 & 3 & LOW & ZERO \\  \hline
         \rowcolor{green!25}
        \textbf{P37} & 206.77 & 65.73 & 40.76 & 22.97 & 35 & 49 & M & N-P-P & 20 & 2058 & 25 & 16 & \cellcolor{red!15}HIGH & HIGH \\  \hline
        \textbf{P38} & 151.15 & 18.15 & 4.09 & 1.68 & 34 & 68 & M & N-N-P & 19 & 2094 & 19 & 11 & \cellcolor{red!15}HIGH & ZERO \\  \hline
        \textbf{P39} & 107.16 & 3.84 & 0.54 & 0 & 66 & 69 & M & N-X-N & 16 & 2052 & 0 & 0 & ZERO & ZERO \\  \hline
        \textbf{P41} & 108.81 & 19.93 & 5.93 & 8.54 & 62 & 62 & M & N-X-P & 16 & 1908 & 29 & 16 & \cellcolor{red!15}HIGH & ZERO \\  \hline
        \textbf{P43} & 102.27 & 5.66 & 1.07 & 1.54 & 60 & 49 & M & N-X-N & 19 & 2014 & 0 & 0 & ZERO & ZERO \\  \hline
         \rowcolor{green!25}
        \textbf{P44} & 157.07 & 34.19 & 22.94 & 15.39 & 32 & 61 & M & N-P-P & 24 & 2055 & 36 & 19 & \cellcolor{red!15}HIGH & HIGH \\  \hline
        \textbf{P45} & 181.73 & 52.19 & 13.30 & 28.55 & 38 & 74 & M & N-X-P & 24 & 1968 & 31 & 15 & \cellcolor{red!15}HIGH & ZERO \\  \hline
         \rowcolor{green!25}
        \textbf{P46} & 149.56 & 15.37 & 2.13 & 3.24 & 39 & 69 & M & P-P-P & 30 & 1962 & 26 & 11 & \cellcolor{red!15}HIGH & HIGH \\  \hline
        \textbf{P48} & 105.46 & 19.93 & 2.1 & 8.91 & 39 & 72 & M & N-N-N & 19 & 1918 & 0 & 0 & ZERO & ZERO \\  \hline
        \textbf{P49} & 87.27 & 7.8 & 0.98 & 0.8 & 49 & 57 & M & N-X-N & 15 & 1698 & 0 & 0 & ZERO & ZERO \\  \hline
         \rowcolor{green!25}
        \textbf{P52} & 97.13 & 13.09 & 4.81 & 5.61 & 35 & 70 & M & N-P-P & 16 & 1972 & 1 & 0 & LOW & HIGH \\  \hline
        \textbf{P53} & 135.17 & 40.98 & 29.08 & 23.37 & 30 & 66 & M & N-X-P & 13 & 1962 & 21 & 11 & \cellcolor{red!15}HIGH & ZERO \\  \hline
        \textbf{P54} & 108.17 & 4.52 & 3.43 & 2.08 & 51 & 27 & M & N-X-N & 20 & 1770 & 0 & 0 & ZERO & ZERO \\  \hline
        \textbf{P55} & 118.09 & 7.46 & 9.57 & 2.34 & 42 & 68 & M & N-X-N & 19 & 2088 & 0 & 0 & ZERO & ZERO \\  \hline
         \rowcolor{green!25}
        \textbf{P56} & 99.52 & 3.45 & 1.32 & 0 & 55 & 50 & M & P-X-N & 19 & 1984 & 0 & 0 & ZERO & LOW \\  \hline
        \textbf{P57} & 79.95 & 7.84 & 0.40 & 0 & 67 & 67 & M & N-X-N & 17 & 1896 & 0 & 0 & ZERO & ZERO \\  \hline
        \textbf{P58} & 81.62 & 0.27 & 0.07 & 0 & 72 & 59 & M & N-X-N & 22 & 1914 & 0 & 0 & ZERO & ZERO \\  \hline
         \rowcolor{green!25}
        \textbf{P5\_P} & 196.49 & 24.31 & 8.22 & 8.01 & 25 & 81 & M & P-P-P & 28 & 1992 & 6 & 4 & LOW & HIGH \\  \hline
        \textbf{P11\_P} & 120.77 & 7.25 & 2.83 & 0 & 54 & 60 & M & N-N-N & 18 & 1962 & 0 & 0 & ZERO & ZERO \\  \hline
         \rowcolor{green!25}
        \textbf{P12\_P} & 121.28 & 21.44 & 8.05 & 13.55 & 37 & 68 & M & P-P-P & 18 & 1998 & 22 & 12 &  \cellcolor{red!15} HIGH & LOW \\  \hline
    \end{tabular}
    \caption{Clinical data (columns 2 to 9 and 15) and simulation results (columns 9 to 14) for the patients included in the study (Green rows: positive clinical cases, Red cells: High risk simulations).
    In column VT, first letter defines if the patient has experienced VT (P) or not (N); second letter is (X) if VT inducibility was not attempted in clinic; if it was attempted, (P) if VT was successfully induced and (N) if not; third letter indicates (P) if any VT was obtained during simulations and (N) otherwise.
    In column `Clinical risk', we present the clinical risk of experiencing VT based on the type of VT that each patient experienced or was induced. We consider VT episodes as HIGH risk when they were identified as sustained monomorphic VT or Polymorphic VT, LOW risk when they were classified as non-sustained VT, and ZERO risk when VT was not experienced or induced. The rest of the columns are explained across the text when referenced.
    }
    \label{tb:datostotal}
\end{table}
\renewcommand{\arraystretch}{1}
\twocolumn


\subsection{Simulation Protocol}\label{sec:stimprotocol}

We have developed a simulation procedure that replicates the extensive pace-mapping protocol used in electrophysiology labs, covering the entire endocardial and epicardial regions of the LV, including the RV septal wall. During the data pre-processing phase, we automatically defined 34 pacing sites, distributed across the 17 AHA segments (17 endocardial sites and 17 epicardial sites). This approach ensures comprehensive simulations from all directions relative to the scar. 
Baseline simulation data (with no additional factors applied) was obtained from an extensive biophysical simulation study that generated restitution curves using the ten Tusscher ionic model~\cite{Tusscher:2004aa}. For this, we employed an S1-S2 protocol with decremental pacing from 295 ms to 270 ms in 5 ms intervals for the S2 stimulus. Initially, a train of six S1 stimuli was delivered with a constant basic cycle length (BCL) of 600 ms, followed by one to three S2 stimuli with a shorter BCL.

To account for variability in the parameters for which we do not have patient-specific data, we have varied independently: i) CV in healthy and BZ tissue, adapting the CV Restitution curves by a factor that ranges between 1 and 1.25; ii) APD restitution curves were varied in a range between 0.75 and 1.25, so that differences at cellular level can also be accounted, see Table~\ref{tb:simprot}. 
By automatically combining all detailed parameters, we simulated more than 3000 scenarios per digital twin to assess the potential risk of a VT episode. We ensured that the variations induced in the different parameters kept the resulting simulations within physiological ranges.


\begin{table}[!h]
    \centering
    \begin{tabular}{l|c|c|c|p{0.4\linewidth}}
        Parameter & Initial & Final & Inc & Description   \\
        \hline
        Ectopic & 1 & 34 & 1 & Location of stimuli (AHA Region) \\
        S2 BCL & 270 & 295 & 5 & BCL (ms) for S2 in S1-S2 protocol \\
        \#S2 & 1 & 3 & 1 & Number of S2 stimuli \\
        CV & 1 & 1.25 & 0.25 & Mult. factor CVR curves \\
        APD & 0.75 & 1.25 & 0.25 & Mult. factor APDR curves\\
    \end{tabular}
    \caption{Simulation Protocol.}
    \label{tb:simprot}
\end{table}

The following parameters were fixed for all simulations: i) frequency of S1 stimuli (600~ms) as in \cite{Arevalo:2016aa}; ii) the number of S1 stimuli (6) for stabilization; iii) weight of CV memory between consecutive stimuli (5\%); and iv) electrotonic effects between neighboring cells (85\%). The total simulation time was also fixed to 6000~ms. Note that the construction and execution of the 3000 simulations per digital twin were conducted automatically and continuously, delivering the results for each patient's scenarios immediately upon completion.


For each case, the following results were recorded: the number of sustained reentries per patient and the associated configurations, including the node ID where the VT initiated, the starting time, the reentry cycle length for at least ten cycles, and the simulation parameters. This allowed us to determine whether positive VT simulations occurred only under extreme parameter values. After the successful completion of the simulations, an automated report was generated for each case, summarizing the results of the simulations.


Given that the majority of patients were on beta-blockers, we reviewed the cases predicted as positive by the previous study which were not induced in the clinic.
For these cases, a second set of 3000 simulations per case was conducted, incorporating the effect of the beta-blockers to determine if the predicted risk changed. Specifically, the values of APD (+45\%) and CV (-6\%) were adjusted based on~\cite{gray2023amiodarone}. Additionally, a clinical follow-up was conducted to determine if any cases had experienced a change in their risk level.

For cases where clinical follow-up was not possible due to the inability to induce VT in the clinic, verification was conducted using biophysical simulations with the cardiac electrophysiology simulator openCARP~\cite{plank2021opencarp} using the ten Tusscher model. Each case was remeshed to meet the model's required resolution with an average of 500$\mu$m compared to Arrhythmic3D's 1000$\mu$m. Fiber orientation was incorporated using a  Laplace-Dirichlet Rule-Based Method, instead of the Streeter model used by Arrhythmic3D. To optimize biophysical simulations, only one reentry per case was simulated to verify if pacing from the same zone induces reentry and to confirm if the exit site corresponds.

To obtain reentry results more quickly, the Rapid Pacing method of openCARP was used. This method involves a train of electrical stimulations with a decreasing coupling interval.
In this case, the coupling interval was set at 10ms. After the last S1,an S2 stimulus was delivered at a frequency of 370ms. If no reentry occurred, the frequency was decreased to 360 ms, and so on, down to 270 ms. If reentry was still not induced, an S3 stimulus was added after each S2, following the entire protocol.



\subsection{Patient Stratification}

Patient stratification was initially explored using current clinical biomarkers alone to establish a baseline ground truth. We analyzed the predictive value of these biomarkers, including LV ejection fraction (LVEF $<$ 35\%), BZ volume~\cite{Roes:2009aa}, scar volume ($>$ 5\% LV) \cite{klem:2012wa} and LV mass \cite{Reinier:2011uv}. Additionally, we incorporated imaging-based metrics related to the BZ region, such as SCC mass, as these have also been reported as predictive of hosting VT isthmuses \cite{Takigawa:2019wh}.

The T-distributed Stochastic Neighbor Embedding (t-SNE) algorithm was employed to visualize the distribution of patients in a 2D space. t-SNE is a widely used non-linear dimensionality reduction algorithm, that  preserves local structure, for visualizing high-dimensional data in lower dimensions, such as two or three dimensions. Subsequently, we constructed a decision tree to assess the complexity of this clinical criteria and quantified the performance of different classification models, including Decision Trees, Random Forests, and Support Vector Machines.

Next, we introduced a new metric, named ARRISK, to assess arrhythmia vulnerability by summarizing the simulation results.
ARRISK is defined through an index called the AR-index, calculated by dividing the number of positive VTs  ($P$) detected during the $N$-simulations performed with a particular digital twin by the total number of effective simulations. Then, the AR-index is  normalized to a range 0-100 to facilitate interpretability.
Once the normalized AR-index was computed, we evaluated its predictive power for clinical diagnosis through a cross-validation process (LOO-nfolds=3) using two classification models: a decision tree and a random forest.
Finally, we define ARRISK as a digital twin of the clinical diagnosis classes (ZERO/LOW/HIGH); ARRISK is classified as ZERO when the patient's AR-index is 0, LOW  when $0<$~AR-i~$\leq~\theta$, and HIGH when AR-i~$>~\theta$, where $\theta$ is a threshold that must be estimated.



\section{Results}

\subsection{Patient-specific simulation studies}
\label{sec:results_sim}

For each of the 51 patients included in the study, a digital twin and computational model were constructed. The set of configurations defined for each digital twin resulted in 3672 3D simulations, which were automatically set up and solved by arrhythmic3D.

From this large set of simulations, some did not complete the stimulation protocol. The simulation studies are designed to stress the tissue and assess the risk of arrhythmia, so some configurations fail when attempting to trigger the S2 stimuli. Since each patient has a unique distribution of healthy, border zone and scar tissue, the properties of the stimulation sites, including cell APD and refractory periods, vary, potentially preventing very short coupling intervals. Additionally, if the stimulus is applied to non-excitable tissue (e.g., the scar core zone, CZ), the protocol fails. However, all cases had over 1500 successful simulations, with most achieving around 2000. The number of simulations for each case is detailed in Table~\ref{tb:datostotal}.



According to the simulation results, the reentry cycle length (CL) of the VTs ranged from 200~ms to 400~ms, shortening as the reentry became established. Consequently, in  positive VT cases, the number of reentry cycles varied between 20 and 30. In our simulation study, 21 patients did not exhibit reentries in any configuration. For cases with reentries, the number of configurations capable of triggering  a VT varied between 1 and 36.

In all cases where reentries were observed, except for P28, at least one positive VT was identified using the default (healthy tissue) parameters. This finding suggests that even patients with minimal tissue electrical remodeling may still be at risk.
Detailed results for all patients, including the columns '\#VT' and '\#VT baseline', are provided in Table~\ref{tb:datostotal}.

\begin{figure}[ht!]
    \begin{center}
    \includegraphics[width=.95\linewidth]{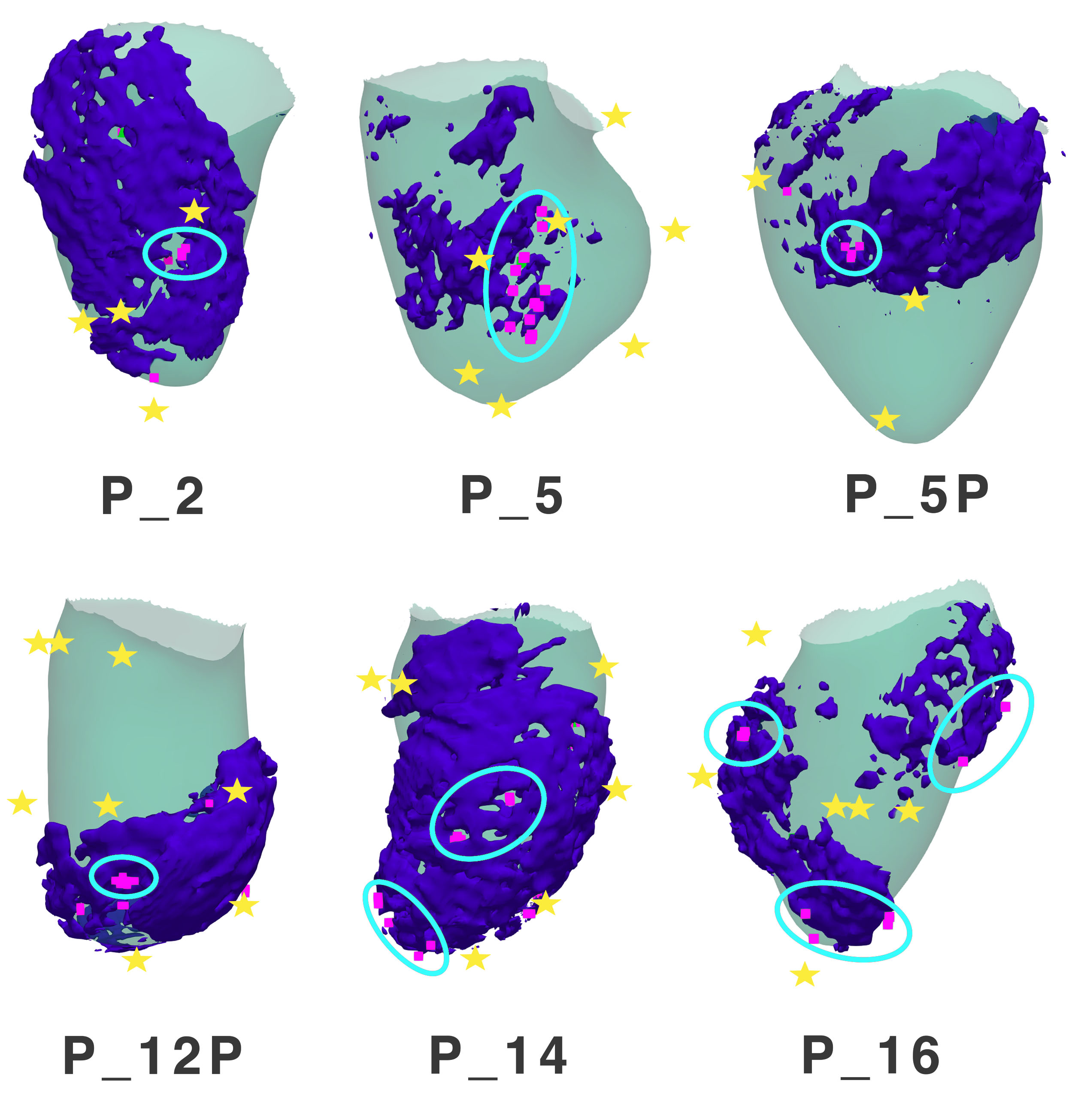} 
    \end{center}
    \caption{Results across multiple example cases, showing all simulations where ventricular tachycardia was induced, visualized on the ventricle of each case. The core zone is depicted in purple. Yellow stars indicate pacing sites, while magenta squares represent the scar exit sites where ventricular tachycardia was initiated. Blue circles highlight areas where exit points frequently cluster around specific hotspots. Despite the reentries being triggered from various pacing sites and parameter configurations, the exit sites consistently cluster in only a few regions.}
    \label{fig:ventClusterSims}
\end{figure}

For intervention planning, identifying the reentry exit site is crucial. The key question is whether the different exit sites observed in a single case converge on a common region of the heart, representing the physiological exit site, or if they are highly dependent on the simulation configuration. Results indicate that, despite the variety of configurations and pacing sites, exit sites consistently appear in just a few regions.
Although the observed VTs may correspond to different circuits, the exit sites  tend to cluster into 2 or 3 large risk zone groups. This is illustrated in Fig.~\ref{fig:ventClusterSims} for 6 patients, where each reentry from the 3000 simulations per case is shown, displaying the exit site nodes (magenta squares) and the initial stimulation nodes (yellow stars), highlighting the clustering of risk zones.
The simulation results also reveal that not all the pacing sites are equally represented in the reentrant configurations. We found that, out of the 34 pacing points distributed across the ventricle, reentries were generated only at sites located in the peri-infarct zone.
%
%
%
%



\begin{figure}[ht!]
    \begin{center}
    \includegraphics[width=\linewidth]{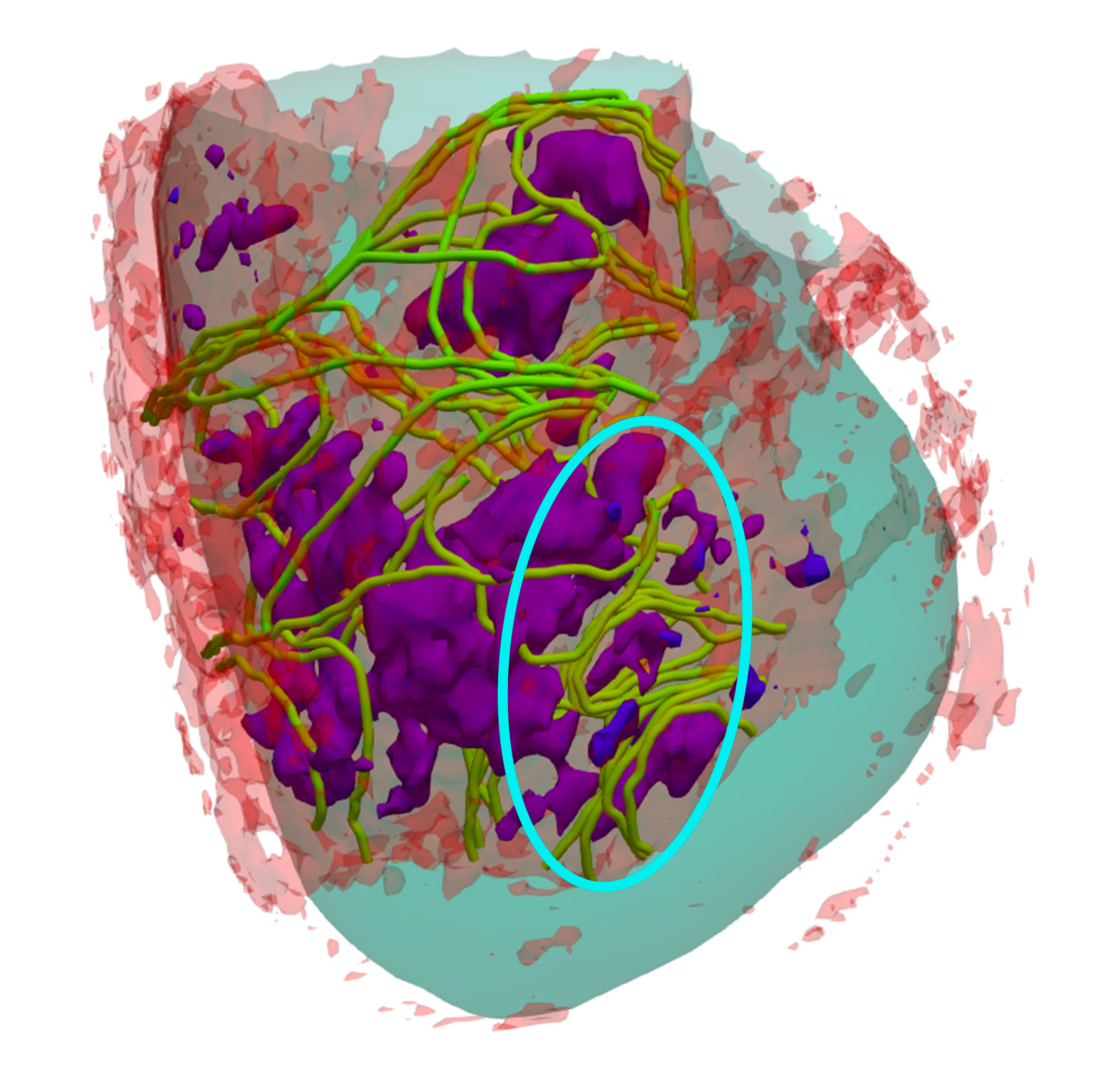} 
    \end{center}
    \caption{
    Detail of the phenotype of patient P5. The border zone is shown in red, the core zone in purple, and the slow conduction channels in green. The blue circle highlights the high-risk slow conduction channels that correspond with the results shown in Fig. ~\ref{fig:ventClusterSims}.}
    \label{fig:caseP5}
\end{figure}

Fig.~\ref{fig:caseP5} provides a closer view of the anatomy of patient P5, highlighting the scar and SCCs.
In this case, which presents a large number of SCCs, only a portion of the 1830 simulations generated reentries, but all were localized within the same risk zone. In the figure, the SCCs located around the exit sites are circled in blue. This patient-specific identification of localized risk zones is clinically relevant, as these areas are potential candidates for ablation.

%
%



\subsection{Risk Stratification}

\begin{figure*}[h]
    \begin{center}
    \includegraphics
        [width=.485\linewidth,
        trim={4.2cm 3.9cm 4.0cm 4.5cm},clip]{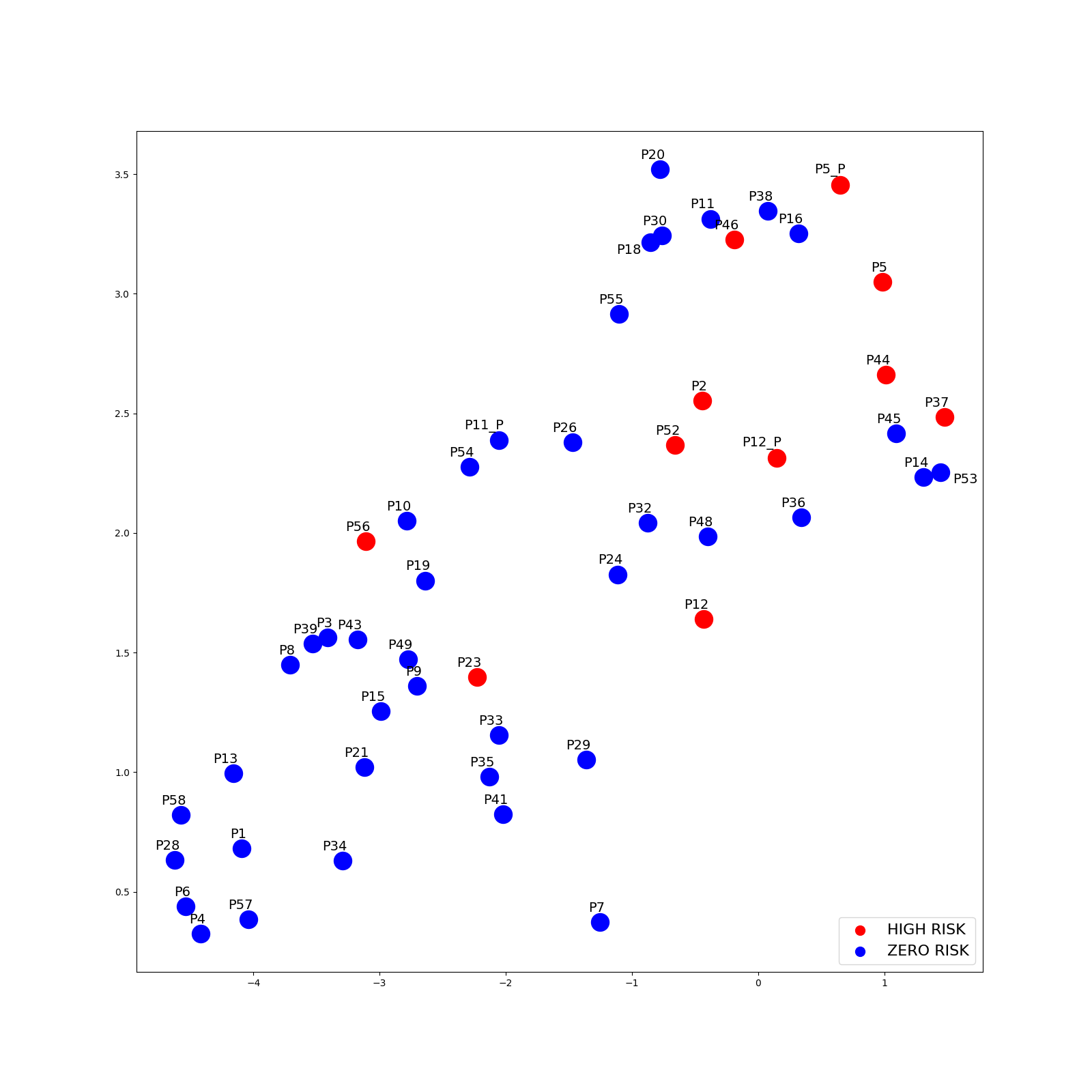}\hfill
    \includegraphics
        [width=.485\linewidth,
        trim={4.2cm 3.9cm 4.0cm 4.5cm},clip]{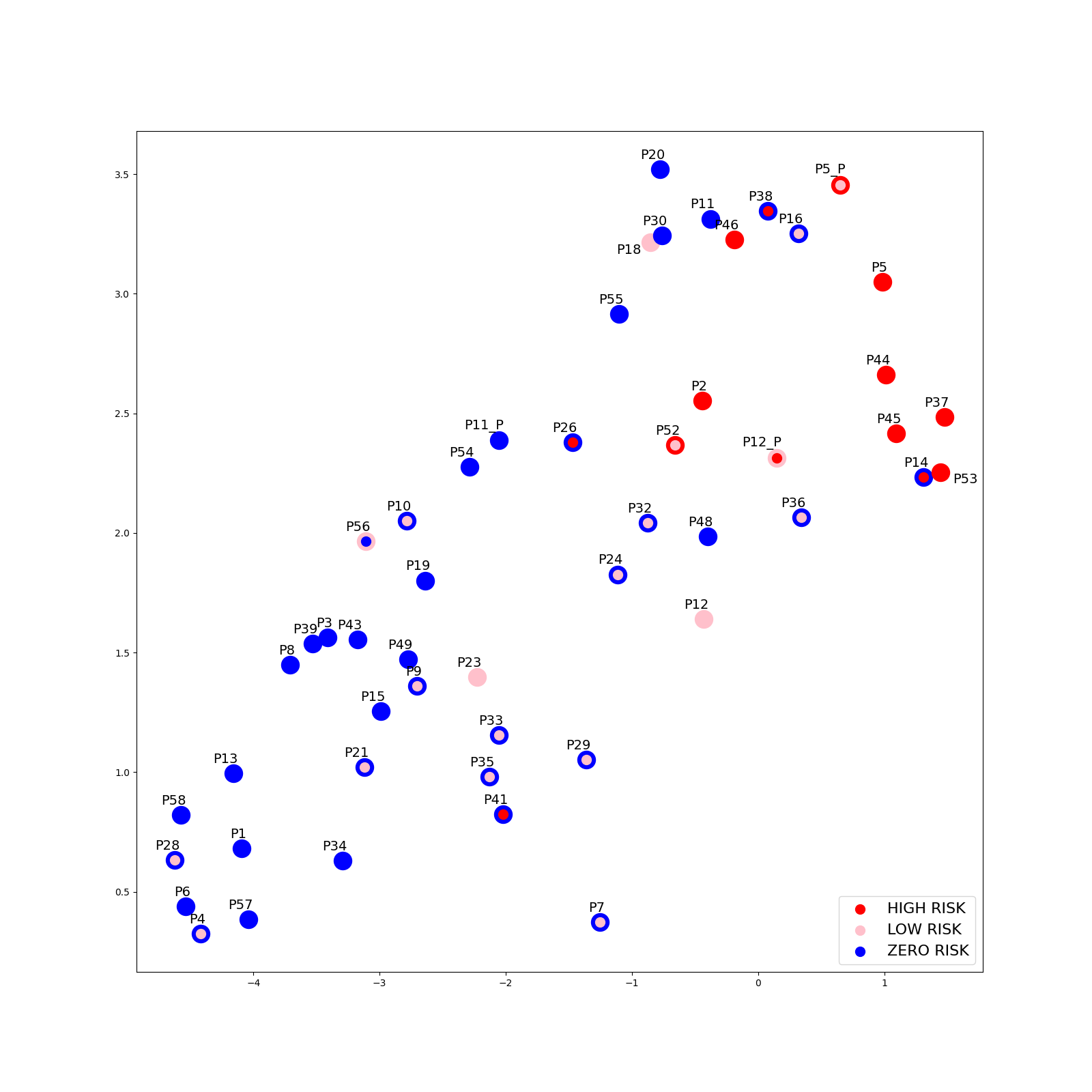}
    \end{center}
    \caption{Left: Visualization of the patient dataset using t-SNE. The color of each point indicates the clinical classification of the patient (red: HIGH risk, blue: ZERO risk). The scale of the axes is derived from the t-SNE projection and does not correspond to the original feature space. Right: ARRISK performance on the 51-case dataset. The outer border represents clinical risk, while the inner circle denotes the ARRISK score (red: HIGH risk, pink: LOW risk, blue: ZERO risk).}
    \label{fig:dataset}
\end{figure*}

Fig.~\ref{fig:dataset} presents the t-SNE analysis for both the ground truth stratification and the ARRISK stratification. It is important to note that t-SNE performs a nonlinear projection from high-dimensional space to low-dimension space (2D in our case) to visually separate clusters. In this projection, the axes do not correspond to any variable in the original data and are unitless.
The left plot shows the analysis for the clinical stratification (ground truth). In this case, positive patients (red color) tend to be cluster in the upper right corner, while the negative patients (blue color) are more dispersed across the space, with a slight tendency to concentrate in the lower left area. Both classes are highly intermixed, indicating that it will be challenging to establish a simple criterion to separate them.

To gain a deeper understanding of the complexity associated with this problem, we also computed the corresponding decision tree, focusing on the two main classes. The results confirm that the dataset is not easily separable, as both branches of the tree contain nodes representing both classes, requiring up to five levels to cover a six-dimensional dataset.
Additionally, we quantified the predictive power of this diagnosis based on clinical data, with the results presented in Table~\ref{table:clasif_res}.
These findings further highlight the challenges of the separability problem, particularly when dealing with an unbalanced dataset with a limited number of samples and features. 
%
\begin{table}[ht!]
\begin{tabular}{@{}llllllllll@{}}
\toprule
 & \multicolumn{2}{l}{Accuracy} & \multicolumn{2}{l}{Precision} & \multicolumn{2}{l}{Recall} & \multicolumn{2}{l}{F1} &  \\ \midrule
          & mean & std  & mean & std  & mean & std & mean & std  &  \\
D.Tree    & 0.64 & 0.09 & 0.25 & 0.17 & 0.36 & 0.3 & 0.28 & 0.09 &  \\
R.Forests & 0.8  & 0.07 & 0.53 & 0.4  & 0.36 & 0.3 & 0.39 & 0.28 &  \\ 
          &      &      &      &      &      &     &      &      &  \\ \bottomrule
\end{tabular}
 \caption{Classification results (cross validation) obtained from the dataset using the clinical criteria}
    \label{table:clasif_res}
    
\end{table}



\begin{figure}[h]
    \centering
    \includegraphics[width=\linewidth]{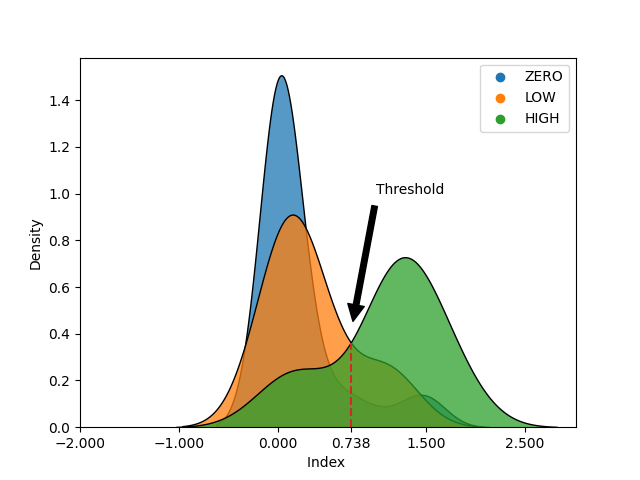}
    \caption{AR-index distributions derived from the clinical diagnosis classes (probability density curves). The value $0.738$ has been selected as the threshold ($\theta$),  required by the ARRISK method to separate the 3 classes defined for this stratification problem.}
    \label{fig:i-distr}
\end{figure}

Next, we present the results obtained from the computation of ARRISK. Fig.~\ref{fig:i-distr} shows the distributions of the AR-index for the three groups of patients based on VT type (ZERO, LOW, HIGH). 
For this dataset, the optimal results are achieved with an AR-index threshold of $\theta=0.738$, which corresponds to the exact point where the distributions of the LOW and HIGH patient groups intersect.
%

\subsection{Clinical Follow-up}
\label{sec:results_followup}

The new ARRISK-based stratification was consistent with the clinical results in the the majority of true (clinical) positive cases.
However, clinically negative cases exhibited different behavior.
At the time of data collection, there were 40 clinically negative cases, and the ARRISK score predicted LOW or HIGH risk for 20 of them.
Moreover, 37 out of the 40 cases had not undergone VTs induction attempts during their treatment at that time. We observed that these patients were on beta-blockers as part of their therapy, which likely caused a mismatch between the patients' clinical conditions and the simulations, as the simulations were conducted in the absence of any treatment.
After updating their clinical history, we noted that three of the cases had subsequently been reported as positive.
Simulations of the remaining 17 clinically negative cases with LOW and HIGH ARRISK scores were repeated following the methodology discussed in Sec.~\ref{sec:stimprotocol}, this time incorporating the effects of beta-blockers.
In Table~\ref{tb:betablock}, it can be observed that after accounting for the effect of beta-blockers, most cases with a LOW ARRISK score shifted to ZERO risk. However, cases with HIGH ARRISK scores continued to present nonzero risk, with the majority still classified as HIGH. All 17 cases were clinically negative, but VT induction was not performed in the clinic, except for case 38, which had a negative outcome. Notably, this case's ARRISK score changed from HIGH to LOW after simulating the effect of beta-blockers.
\begin{table}[!ht]
    \centering
    \begin{tabular}{ccc >{\centering\arraybackslash} p{0.2\linewidth}}
    \hline
        \textbf{PATIENT} & \textbf{Clinical Risk} & \textbf{ARRISK} & \textbf{ARRISK with beta-blockers} \\ \hline
        \textbf{P4} & ZERO & LOW & ZERO \\
        \rowcolor{green!25}
        \textbf{P7} & ZERO & LOW & LOW \\ 
        \textbf{P9} & ZERO & LOW & ZERO \\ 
        \textbf{P10} & ZERO & LOW & ZERO \\ 
        \rowcolor{green!25}
        \textbf{P14} & ZERO & HIGH & HIGH \\ 
        \rowcolor{green!25}
        \textbf{P16} & ZERO & LOW & LOW \\ 
        \textbf{P21} & ZERO & LOW & ZERO \\ 
        \textbf{P24} & ZERO & LOW & ZERO \\ 
        \rowcolor{green!25}
        \textbf{P26} & ZERO & HIGH & HIGH \\ 
        \textbf{P28} & ZERO & LOW & ZERO \\ 
        \textbf{P29} & ZERO & LOW & ZERO \\ 
        \textbf{P32} & ZERO & LOW & ZERO \\ 
        \textbf{P33} & ZERO & LOW & ZERO \\ 
        \textbf{P35} & ZERO & LOW & ZERO \\ 
        \textbf{P36} & ZERO & LOW & ZERO \\ 
        \rowcolor{green!25}
        \textbf{P38} & ZERO & HIGH & LOW \\ 
        \rowcolor{green!25}
        \textbf{P41} & ZERO & HIGH & HIGH \\ \hline
    \end{tabular}
    \caption{Simulations with the effect of beta-blockers, in clinically negative cases with positive results in the initial simulations of the study. Rows shaded in greed correspond to positive cases with beta-blocker effect. None of the patients in this table were induced VT in the clinic except for case P38, which gave negative result. Most of the patients with LOW ARRISK score changed to ZERO after simulating the effect of beta-blockers.}
    \label{tb:betablock}
\end{table}

Fig.~\ref{fig:dataset}, right, shows the t-SNE projection of the patients, focusing on their positive and negative classification according to ARRISK, and comparing it to the clinical risk outcome. The color in the center of each point represents the ARRISK class, while the border color corresponds to the clinical class of the dataset. 
The three clinical cases that changed from negative to positive labels after follow-up have been updated in this graph compared to Fig.~\ref{fig:dataset}, left.
Among the clinically positive cases (red and pink), ARRISK coincides in 13 out of 14 cases, with the risk level matching in 77\% of these 13 cases. Only one clinically positive case with LOW risk is classified as negative. Regarding negative cases (blue), ARRISK matches the clinical result in 20 out of 37 cases, distributing the risk among the 17 non-matching cases, with 13 classified as LOW risk and 4 as HIGH risk.

Therefore, only 4 out of the 17 cases with HIGH risk should be considered for potential ablation. These 4 cases can be identified in Fig. \ref{fig:dataset} by a blue border with a red inner circle. Case P14 is positioned near positive neighboring cases, whereas cases P41, P38, and P26 are situated closer to negative neighbors. Consequently, these cases present greater difficulty in stratification based solely on pre-simulation data.

%
In Table~\ref{tb:opencarpSelect}, we present the  Arrhythmic3D results for these HIGH-risk cases. The table specifies the AHA segment from which pacing was applied (Pacing site), the AHA segment where the reentry initiated (Exit site), and the Cycle Length of the reentry. All cases correspond to sustained ventricular tachycardia.
\begin{table}[!ht]
    \centering
    \begin{tabular}{cccc}
    \hline
        \multicolumn{4}{c}{\textbf{Arrhythmic3D}} \\ \hline
        \textbf{Case} & AHA stim & AHA exit & CL \\ 
        \textbf{P14} & 13 & 7/13 & S4 - 295 \\ 
        \textbf{P26} & 7 & 12 & S4 - 290 \\ 
        \textbf{P38} & 9 & 9 & S4 - 280 \\ 
        \textbf{P41} & 7 & 8 & S4 - 290 \\ \hline
    \end{tabular}
    \caption{Reentry information for 4 HIGH-risk cases selected for further verification of Pacing site and Exit site using openCARP. The information in this table corresponds to the results obtained with Arrhythmic3D. All reentries correspond to Sustained Ventricular Tachycardia.  AHA stim: AHA segment used for stimulation; AHA exit: AHA segment where exit took place; CL: Cycle Length Reentry. }
    \label{tb:opencarpSelect}
\end{table}

For the verification of these 4 HIGH-risk cases, biophysical simulations were conducted using the cardiac electrophysiology simulator openCARP. Table~\ref{tb:opencarpResults}, presents the results obtained with openCARP. Reentries were successfully induced in all 4 cases by stimulating from the same pacing zone, and the initiation site of the reentry was consistent across cases. Additionally, the cycle length of the reentries showed minor differences, with variations of up to 10 ms at most. In all cases, the outcome was also a sustained ventricular tachycardia, in agreement with Arrhythmic3D results. These findings confirm the accuracy of the reentries obtained and the risk zones detected by Arrhythmic3D. 
\begin{table}[!ht]
    \centering
    \begin{tabular}{cccc}
    \hline
        \multicolumn{4}{c}{\textbf{Biophysical Simulations - openCARP}} \\ \hline
        \textbf{Case} & AHA stim & AHA exit & CL \\ 
        \textbf{P14} & 13 & 7/13 & S2 - 310 \\ 
        \textbf{P26} & 7 & 12 & S2 - 310 \\ 
        \textbf{P38} & 9 & 9 & S2 - 280 \\ 
        \textbf{P41} & 7 & 8 & S2 - 290 \\ \hline
    \end{tabular}
    \caption{Results of in silico pacing using openCARP biophysical solver, using the same pacing site used with Arrhythmic3D. As in Table~\ref{tb:opencarpSelect}, all cases present Sustained Ventricular Tachycardia. AHA stim: AHA segment used for stimulation; AHA exit: AHA segment where exit took place; CL: Cycle Length Reentry.}
    \label{tb:opencarpResults}
\end{table}

\begin{figure}[h]
    \begin{center}
    \includegraphics[width=.95\linewidth]{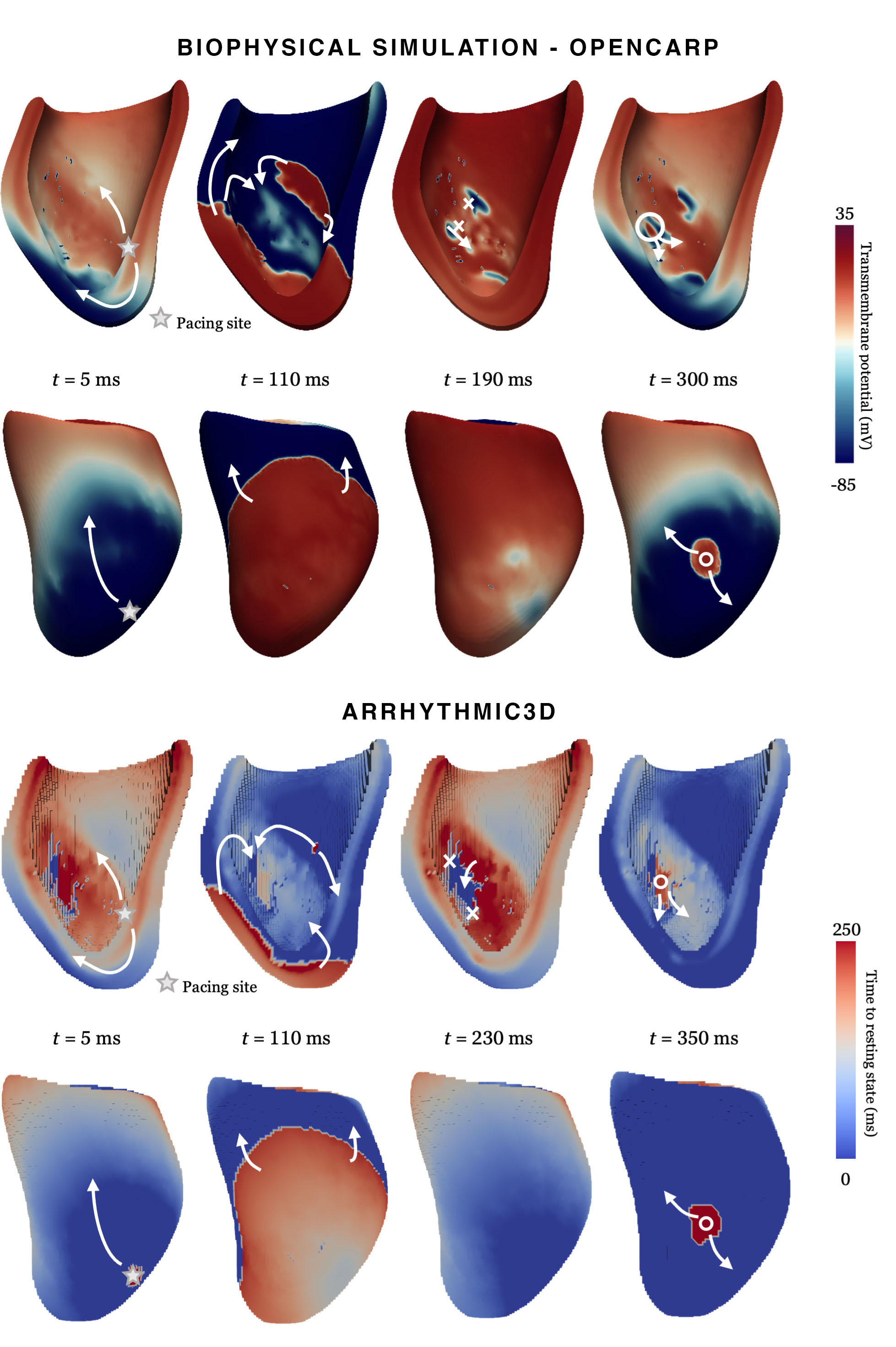} 
    \end{center}
    \caption{Comparison of ventricular tachycardia initiation in case P41. (Top) Biophysical simulation with openCARP, (Bottom) Simulation with Arrhythmic3D. In each simulation, the snapshots correspond to the posterior view (top) and the anterior view (bottom) of the left ventricle. The time instant above each map is measured from the delivery of the last pacing stimulus. White arrows indicate the direction of propagation, white crosses mark propagation blocks. The gray star denotes the initial pacing site, and the white circle represents the initial exit site of the reentry.}
    \label{fig:opencarp_vs_CA}
\end{figure}

We find good agreement between clinical results and the ARRISK score where the clinical classification is more conclusive, while cases where ARRISK might be considered to fail could still have their classification in clinical history. Moreover, among the 17 non-matching negative cases, ARRISK returned 13 LOW-risk scores, and 11 of these changed to ZERO when beta-blockers were simulated.
For further insight, Fig.~\ref{fig:opencarp_vs_CA} presents a comparison between the biophysical simulations with openCARP and Arrhythmic3D for one of the reentries obtained from case P41, highlighting the previously noted similarities.

\section{Discussion}

The results presented in Fig.~\ref{fig:dataset} demonstrate a strong concordance between the proposed ARRISK score and clinical labeling, especially in positive cases—HIGH risk (red) and LOW risk (pink)—and negative cases. However, notable discrepancies were observed in negative clinical cases, where several LOW and HIGH-risk scores were assigned by ARRISK. Interestingly, only in three of these cases (P38, P48, $P11_P$) were attempts made to induce VTs clinically. For these three cases, ARRISK aligned with the negative result in P48 and $P11_P$, where no reentries were induced by septal wall stimulation, as performed by the medical team. In case P38, despite negative clinical inducibility, our simulations still indicated a reentry risk, as discussed in Sec.~\ref{sec:results_followup}.

For the remaining cases, no inducibility test was conducted, and these patients were under beta-blocker therapy, which may have influenced their clinical status. The absence of clinical confirmation of zero risk suggests that their classification as negative could be provisional. This observation is supported by prior studies, such as those by Zhou et al. \cite{Zhou:2021ua} and Trayanova et al. \cite{trayanova2022deep}, which emphasize the potential of computational models to uncover latent arrhythmic risks not detectable by standard clinical procedures. The possibility of disease progression and eventual VT manifestation in these patients underscores the utility of ARRISK in providing an early warning where clinical assessments might fall short.
Comparing our findings with the work of Aronis et al. \cite{aronis2021characterization}, who demonstrated the power of patient-specific models in predicting arrhythmic events, we see a consistent pattern where computational models provide insights beyond traditional imaging and electrophysiological tests. In particular, Aronis et al. highlighted that computational models could identify high-risk myocardial regions not apparent through standard imaging. Our study similarly found that ARRISK was able to detect potential reentry risks in cases where clinical tests failed to induce VTs, reinforcing the value of combining computational and clinical approaches.

The ability to perform a large number of simulations with varied parameters and model configurations allowed us to assess the stability of the ARRISK score. As demonstrated, expanding the range of restitution curve values derived from the ten Tusscher model did not significantly alter patient stratification in most cases. Instead, simulations with varied APD and CV values often confirmed the same reentries despite slight parameter variations. This is critical for identifying highly proarrhythmic cases. This approach is consistent with the findings of Deng et al. \cite{Deng:2019aa} and Prakosa et al. \cite{Prakosa:2018aa}, who noted that regions with dense scar tissue and slow conduction channels are more likely to sustain reentrant circuits, particularly when these regions consistently produce reentries under varied simulation conditions.

The robustness of ARRISK is further evidenced by our observation that if a scar area can reproduce the same reentry with different simulation parameters, it is more likely to generate reentries compared to scars that produce reentries only under a specific set of conditions. This insight is consistent with the work of Arevalo et al. \cite{Arevalo:2016aa}, who developed the VARP index to assess the risk of sudden cardiac death (SCD) using multi-site pacing and biophysical simulations. They found that patients whose models could trigger VT from multiple pacing locations were at a higher risk of SCD. Our study extends this concept by incorporating a larger number of pacing sites (34 compared to 19 in Arevalo et al.'s study) and by varying additional tissue properties such as CVR and APDR, further enhancing the predictive accuracy of the model.

Our findings align with Sung et al. \cite{sung2020personalized}, \cite{sung2023wavefront}, who emphasized the importance of the interaction between scar morphology and conduction properties in arrhythmia formation. Our results confirm that SCCs significantly contribute to the generation and maintenance of VTs. However, we observed that not all large SCC masses are associated with a high risk of reentry, highlighting the need to consider the specific morphological relationship between the BZ and CZ. This nuance was also noted by Sung et al., who pointed out that complex scar geometries, especially those involving intricate SCC networks, require detailed analysis to accurately assess arrhythmic risk.

Moreover, our methodology, which allows for the simulation of thousands of scenarios within a single day, represents a significant advancement toward bridging the gap between in-silico studies and clinical practice. As Xie et al. \cite{xie2022advanced} argued, computational models like ARRISK could be integrated into clinical workflows to provide real-time risk assessments, thereby enabling more timely and targeted interventions. Our system's ability to provide a comprehensive risk assessment, including the identification of potential ablation sites, within such a short time frame underscores its practical applicability in clinical settings.

The validation of our results through biophysical simulations using openCARP further strengthens the credibility of our approach. Okada et al. \cite{okada2020substrate} highlighted the importance of validating computational predictions with detailed biophysical models to ensure their clinical relevance. In our study, the agreement between the reentries and risk zones detected by ARRISK and those confirmed by openCARP simulations demonstrates the reliability of our method. This validation step is crucial for ensuring that the insights gained from ARRISK are not only theoretically sound but also applicable in real-world clinical scenarios.

In summary, the ARRISK score, supported by extensive computational simulations and validated through biophysical modeling, offers a promising tool for improving the stratification and management of patients at risk of ventricular tachycardia. By integrating advanced computational models with clinical data, we can enhance the accuracy of arrhythmic risk predictions, leading to better-informed decisions regarding patient management. This approach aligns with the broader trend in precision medicine, where tailored treatments based on individual risk profiles are increasingly recognized as the key to improving outcomes in complex cardiac conditions.

\section{Conclusions}
In this study, we introduced the ARRISK score, a computational tool designed to enhance the stratification and management of patients at risk of ventricular tachycardia (VT). Through extensive simulations incorporating a wide range of electrophysiological parameters, ARRISK demonstrated strong concordance with clinical outcomes, particularly in identifying high-risk patients. Our findings underscore the value of integrating computational models with clinical data to provide a more comprehensive assessment of arrhythmic risk. The validation of our approach through biophysical simulations further supports the clinical applicability of ARRISK. This tool represents a significant step toward more personalized and precise treatment strategies in cardiac care, ultimately contributing to improved patient outcomes.


\bibliographystyle{IEEEtran}
\bibliography{tmi24}

\begin{thebibliography}{10}
\providecommand{\url}[1]{#1}
\csname url@samestyle\endcsname
\providecommand{\newblock}{\relax}
\providecommand{\bibinfo}[2]{#2}
\providecommand{\BIBentrySTDinterwordspacing}{\spaceskip=0pt\relax}
\providecommand{\BIBentryALTinterwordstretchfactor}{4}
\providecommand{\BIBentryALTinterwordspacing}{\spaceskip=\fontdimen2\font plus
\BIBentryALTinterwordstretchfactor\fontdimen3\font minus
  \fontdimen4\font\relax}
\providecommand{\BIBforeignlanguage}[2]{{%
\expandafter\ifx\csname l@#1\endcsname\relax
\typeout{** WARNING: IEEEtran.bst: No hyphenation pattern has been}%
\typeout{** loaded for the language `#1'. Using the pattern for}%
\typeout{** the default language instead.}%
\else
\language=\csname l@#1\endcsname
\fi
#2}}
\providecommand{\BIBdecl}{\relax}
\BIBdecl

\bibitem{Pouleur:2010aa}
A.-C. Pouleur, E.~Barkoudah, H.~Uno, H.~Skali, P.~V. Finn, S.~L. Zelenkofske,
  Y.~N. Belenkov, V.~Mareev, E.~J. Velazquez, J.~L. Rouleau, A.~P. Maggioni,
  L.~K{\o}ber, R.~M. Califf, J.~J.~V. McMurray, M.~A. Pfeffer, S.~D. Solomon,
  and {VALIANT Investigators}, ``Pathogenesis of sudden unexpected death in a
  clinical trial of patients with myocardial infarction and left ventricular
  dysfunction, heart failure, or both,'' \emph{Circulation}, vol. 122, no.~6,
  pp. 597--602, Aug 2010.

\bibitem{Donahue:2024aa}
J.~K. Donahue, J.~Chrispin, and O.~A. Ajijola, ``Mechanism of ventricular
  tachycardia occurring in chronic myocardial infarction scar,'' \emph{Circ
  Res}, vol. 134, no.~3, pp. 328--342, Feb 2024.

\bibitem{stevenson1998radiofrequency}
W.~G. Stevenson, P.~L. Friedman, D.~Kocovic, P.~T. Sager, L.~A. Saxon, and
  B.~Pavri, ``Radiofrequency catheter ablation of ventricular tachycardia after
  myocardial infarction,'' \emph{Circulation}, vol.~98, no.~4, pp. 308--314,
  1998.

\bibitem{Cronin:2020aa}
E.~M. Cronin, F.~M. Bogun, P.~Maury, P.~Peichl, M.~Chen, N.~Namboodiri,
  L.~Aguinaga, L.~R. Leite, S.~M. Al-Khatib, E.~Anter, A.~Berruezo, D.~J.
  Callans, M.~K. Chung, P.~Cuculich, A.~d'Avila, B.~J. Deal, P.~D. Bella,
  T.~Deneke, T.-M. Dickfeld, C.~Hadid, H.~M. Haqqani, G.~N. Kay,
  R.~Latchamsetty, F.~Marchlinski, J.~M. Miller, A.~Nogami, A.~R. Patel, R.~K.
  Pathak, L.~C. Saenz~Morales, P.~Santangeli, J.~L. Sapp, Jr, A.~Sarkozy,
  K.~Soejima, W.~G. Stevenson, U.~B. Tedrow, W.~S. Tzou, N.~Varma, and
  K.~Zeppenfeld, ``2019 hrs/ehra/aphrs/lahrs expert consensus statement on
  catheter ablation of ventricular arrhythmias: Executive summary,'' \emph{J
  Arrhythm}, vol.~36, no.~1, pp. 1--58, Feb 2020.

\bibitem{Trayanova:2024aa}
N.~A. Trayanova and A.~Prakosa, ``Up digital and personal: How heart digital
  twins can transform heart patient care,'' \emph{Heart Rhythm}, vol.~21,
  no.~1, pp. 89--99, Jan 2024.

\bibitem{Corral-Acero:2020aa}
J.~Corral-Acero, F.~Margara, M.~Marciniak, C.~Rodero, F.~Loncaric, Y.~Feng,
  A.~Gilbert, J.~F. Fernandes, H.~A. Bukhari, A.~Wajdan, M.~V. Martinez, M.~S.
  Santos, M.~Shamohammdi, H.~Luo, P.~Westphal, P.~Leeson, P.~DiAchille,
  V.~Gurev, M.~Mayr, L.~Geris, P.~Pathmanathan, T.~Morrison, R.~Cornelussen,
  F.~Prinzen, T.~Delhaas, A.~Doltra, M.~Sitges, E.~J. Vigmond, E.~Zacur,
  V.~Grau, B.~Rodriguez, E.~W. Remme, S.~Niederer, P.~Mortier, K.~McLeod,
  M.~Potse, E.~Pueyo, A.~Bueno-Orovio, and P.~Lamata, ``The 'digital twin' to
  enable the vision of precision cardiology,'' \emph{Eur Heart J}, vol.~41,
  no.~48, pp. 4556--4564, Dec 2020.

\bibitem{ten2007organization}
K.~H. Ten~Tusscher, R.~Hren, and A.~V. Panfilov, ``Organization of ventricular
  fibrillation in the human heart,'' \emph{Circulation research}, vol. 100,
  no.~12, pp. e87--e101, 2007.

\bibitem{Niederer:2011wg}
S.~A. Niederer, E.~Kerfoot, A.~P. Benson, M.~O. Bernabeu, O.~Bernus,
  C.~Bradley, E.~M. Cherry, R.~Clayton, F.~H. Fenton, A.~Garny, E.~Heidenreich,
  S.~Land, M.~Maleckar, P.~Pathmanathan, G.~Plank, J.~F. Rodr{\'\i}guez,
  I.~Roy, F.~B. Sachse, G.~Seemann, O.~Skavhaug, and N.~P. Smith,
  ``Verification of cardiac tissue electrophysiology simulators using an
  n-version benchmark,'' \emph{Philos Trans A Math Phys Eng Sci}, vol. 369, no.
  1954, pp. 4331--51, Nov 2011.

\bibitem{Chen:2016aa}
Z.~Chen, R.~Cabrera-Lozoya, J.~Relan, M.~Sohal, A.~Shetty, R.~Karim,
  H.~Delingette, J.~Gill, K.~Rhode, N.~Ayache, P.~Taggart, C.~A. Rinaldi,
  M.~Sermesant, and R.~Razavi, ``Biophysical modeling predicts ventricular
  tachycardia inducibility and circuit morphology: A combined clinical
  validation and computer modeling approach,'' \emph{J Cardiovasc
  Electrophysiol}, vol.~27, no.~7, pp. 851--60, 07 2016.

\bibitem{Deng:2019aa}
D.~Deng, A.~Prakosa, J.~Shade, P.~Nikolov, and N.~A. Trayanova,
  ``Characterizing conduction channels in postinfarction patients using a
  personalized virtual heart,'' \emph{Biophys J}, vol. 117, no.~12, pp.
  2287--2294, 12 2019.

\bibitem{Lopez-Perez:2019aa}
A.~Lopez-Perez, R.~Sebastian, M.~Izquierdo, R.~Ruiz, M.~Bishop, and J.~M.
  Ferrero, ``Personalized cardiac computational models: From clinical data to
  simulation of infarct-related ventricular tachycardia,'' \emph{Front
  Physiol}, vol.~10, p. 580, 2019.

\bibitem{Prakosa:2018aa}
A.~Prakosa, H.~J. Arevalo, D.~Deng, P.~M. Boyle, P.~P. Nikolov, H.~Ashikaga,
  J.~J.~E. Blauer, E.~Ghafoori, C.~J. Park, R.~C. Blake, 3rd, F.~T. Han, R.~S.
  MacLeod, H.~R. Halperin, D.~J. Callans, R.~Ranjan, J.~Chrispin, S.~Nazarian,
  and N.~A. Trayanova, ``Personalized virtual-heart technology for guiding the
  ablation of infarct-related ventricular tachycardia,'' \emph{Nat Biomed Eng},
  vol.~2, no.~10, pp. 732--740, Oct 2018.

\bibitem{Zhang:2023aa}
Y.~Zhang, K.~Zhang, A.~Prakosa, C.~James, S.~L. Zimmerman, R.~Carrick, E.~Sung,
  A.~Gasperetti, C.~Tichnell, B.~Murray, H.~Calkins, and N.~A. Trayanova,
  ``Predicting ventricular tachycardia circuits in patients with arrhythmogenic
  right ventricular cardiomyopathy using genotype-specific heart digital
  twins,'' \emph{Elife}, vol.~12, Oct 2023.

\bibitem{Bhagirath:2023aa}
P.~Bhagirath, F.~O. Campos, P.~Postema, M.~J.~B. Kemme, A.~A.~M. Wilde, A.~J.
  Prassl, A.~Neic, C.~A. Rinaldi, M.~J.~W. G{\"o}tte, G.~Plank, and M.~J.
  Bishop, ``Arrhythmogenic vulnerability of re-entrant pathways in post-infarct
  ventricular tachycardia assessed by advanced computational modelling,''
  \emph{Europace}, vol.~25, no.~9, Aug 2023.

\bibitem{Loewe:2018uz}
A.~Loewe, E.~Poremba, T.~Oesterlein, A.~Luik, C.~Schmitt, G.~Seemann, and
  O.~D{\"o}ssel, ``Patient-specific identification of atrial flutter
  vulnerability-a computational approach to reveal latent reentry pathways,''
  \emph{Front Physiol}, vol.~9, p. 1910, 2018.

\bibitem{Neic:2017um}
A.~Neic, F.~O. Campos, A.~J. Prassl, S.~A. Niederer, M.~J. Bishop, E.~J.
  Vigmond, and G.~Plank, ``Efficient computation of electrograms and ecgs in
  human whole heart simulations using a reaction-eikonal model,'' \emph{J
  Comput Phys}, vol. 346, pp. 191--211, Oct 2017.

\bibitem{serra:2022au}
D.~Serra, P.~Romero, I.~Garcia-Fernandez, M.~Lozano, A.~Liberos, M.~Rodrigo,
  A.~Bueno-Orovio, A.~Berruezo, and R.~Sebastian, ``An automata-based cardiac
  electrophysiology simulator to assess arrhythmia inducibility,''
  \emph{Mathematics}, vol.~10, no.~8, p. 1293, 2022.

\bibitem{Mendoca:2021aa}
C.~Mendonca~Costa, P.~Gemmell, M.~K. Elliott, J.~Whitaker, F.~O. Campos,
  M.~Strocchi, A.~Neic, K.~Gillette, E.~Vigmond, G.~Plank, R.~Razavi,
  M.~O'Neill, C.~A. Rinaldi, and M.~J. Bishop, ``Determining anatomical and
  electrophysiological detail requirements for computational ventricular models
  of porcine myocardial infarction,'' \emph{Comput Biol Med}, vol. 141, p.
  105061, Nov 2021.

\bibitem{Lopez-Perez:2015aa}
A.~Lopez-Perez, R.~Sebastian, and J.~M. Ferrero, ``Three-dimensional cardiac
  computational modelling: methods, features and applications,'' \emph{Biomed
  Eng Online}, vol.~14, p.~35, Apr 2015.

\bibitem{streeter1969fiber}
D.~D. Streeter~Jr, H.~M. Spotnitz, D.~P. Patel, J.~Ross~Jr, and E.~H.
  Sonnenblick, ``Fiber orientation in the canine left ventricle during diastole
  and systole,'' \emph{Circulation research}, vol.~24, no.~3, pp. 339--347,
  1969.

\bibitem{Tusscher:2004aa}
K.~H. W.~J. ten Tusscher, D.~Noble, P.~J. Noble, and A.~V. Panfilov, ``A model
  for human ventricular tissue,'' \emph{Am J Physiol Heart Circ Physiol}, vol.
  286, no.~4, pp. H1573--89, Apr 2004.

\bibitem{Arevalo:2016aa}
H.~J. Arevalo, F.~Vadakkumpadan, E.~Guallar, A.~Jebb, P.~Malamas, K.~C. Wu, and
  N.~A. Trayanova, ``Arrhythmia risk stratification of patients after
  myocardial infarction using personalized heart models,'' \emph{Nat Commun},
  vol.~7, p. 11437, 05 2016.

\bibitem{gray2023amiodarone}
R.~A. Gray and M.~R. Franz, ``Amiodarone prevents wave front-tail interactions
  in patients with heart failure: an in silico study,'' \emph{American Journal
  of Physiology-Heart and Circulatory Physiology}, vol. 325, no.~5, pp.
  H952--H964, 2023.

\bibitem{plank2021opencarp}
G.~Plank, A.~Loewe, A.~Neic, C.~Augustin, Y.-L. Huang, M.~A. Gsell,
  E.~Karabelas, M.~Nothstein, A.~J. Prassl, J.~S{\'a}nchez \emph{et~al.}, ``The
  opencarp simulation environment for cardiac electrophysiology,''
  \emph{Computer methods and Programs in Biomedicine}, vol. 208, p. 106223,
  2021.

\bibitem{Roes:2009aa}
S.~D. Roes, C.~J.~W. Borleffs, R.~J. van~der Geest, J.~J.~M. Westenberg, N.~A.
  Marsan, T.~A.~M. Kaandorp, J.~H.~C. Reiber, K.~Zeppenfeld, H.~J. Lamb,
  A.~de~Roos, M.~J. Schalij, and J.~J. Bax, ``Infarct tissue heterogeneity
  assessed with contrast-enhanced mri predicts spontaneous ventricular
  arrhythmia in patients with ischemic cardiomyopathy and implantable
  cardioverter-defibrillator,'' \emph{Circ Cardiovasc Imaging}, vol.~2, no.~3,
  pp. 183--90, May 2009.

\bibitem{klem:2012wa}
I.~Klem, J.~W. Weinsaft, T.~D. Bahnson, D.~Hegland, H.~W. Kim, B.~Hayes, M.~A.
  Parker, R.~M. Judd, and R.~J. Kim, ``Assessment of myocardial scarring
  improves risk stratification in patients evaluated for cardiac defibrillator
  implantation,'' \emph{J Am Coll Cardiol}, vol.~60, no.~5, pp. 408--20, Jul
  2012.

\bibitem{Reinier:2011uv}
K.~Reinier, C.~Dervan, T.~Singh, A.~Uy-Evanado, S.~Lai, K.~Gunson, J.~Jui, and
  S.~S. Chugh, ``Increased left ventricular mass and decreased left ventricular
  systolic function have independent pathways to ventricular arrhythmogenesis
  in coronary artery disease,'' \emph{Heart Rhythm}, vol.~8, no.~8, pp.
  1177--82, Aug 2011.

\bibitem{Takigawa:2019wh}
M.~Takigawa, J.~Duchateau, F.~Sacher, R.~Martin, K.~Vlachos, T.~Kitamura,
  M.~Sermesant, N.~Cedilnik, G.~Cheniti, A.~Frontera, N.~Thompson, C.~Martin,
  G.~Massoullie, F.~Bourier, A.~Lam, M.~Wolf, W.~Escande, C.~Andr{\'e},
  T.~Pambrun, A.~Denis, N.~Derval, M.~Hocini, M.~Haissaguerre, H.~Cochet, and
  P.~Ja{\"\i}s, ``Are wall thickness channels defined by computed tomography
  predictive of isthmuses of postinfarction ventricular tachycardia?''
  \emph{Heart Rhythm}, vol.~16, no.~11, pp. 1661--1668, 11 2019.

\bibitem{Zhou:2021ua}
S.~Zhou, A.~AbdelWahab, J.~L. Sapp, E.~Sung, K.~N. Aronis, J.~W. Warren, P.~J.
  MacInnis, R.~Shah, B.~M. Hor{\'a}{\v c}ek, R.~Berger, H.~Tandri, N.~A.
  Trayanova, and J.~Chrispin, ``Prospective multicenter assessment of a new
  intraprocedural automated system for localizing idiopathic ventricular
  arrhythmia origins,'' \emph{JACC Clin Electrophysiol}, vol.~7, no.~3, pp.
  395--407, 03 2021.

\bibitem{trayanova2022deep}
N.~A. Trayanova and E.~J. Topol, ``Deep learning a person's risk of sudden
  cardiac death,'' \emph{The Lancet}, vol. 399, no. 10339, p. 1933, 2022.

\bibitem{aronis2021characterization}
K.~N. Aronis, A.~Prakosa, T.~Bergamaschi, R.~D. Berger, P.~M. Boyle,
  J.~Chrispin, S.~Ju, J.~E. Marine, S.~Sinha, H.~Tandri \emph{et~al.},
  ``Characterization of the electrophysiologic remodeling of patients with
  ischemic cardiomyopathy by clinical measurements and computer simulations
  coupled with machine learning,'' \emph{Frontiers in Physiology}, vol.~12, p.
  684149, 2021.

\bibitem{sung2020personalized}
E.~Sung, A.~Prakosa, K.~N. Aronis, S.~Zhou, S.~L. Zimmerman, H.~Tandri,
  S.~Nazarian, R.~D. Berger, J.~Chrispin, and N.~A. Trayanova, ``Personalized
  digital-heart technology for ventricular tachycardia ablation targeting in
  hearts with infiltrating adiposity,'' \emph{Circulation: Arrhythmia and
  Electrophysiology}, vol.~13, no.~12, p. e008912, 2020.

\bibitem{sung2023wavefront}
E.~Sung, A.~Prakosa, S.~Kyranakis, R.~D. Berger, J.~Chrispin, and N.~A.
  Trayanova, ``Wavefront directionality and decremental stimuli synergistically
  improve identification of ventricular tachycardia substrate: insights from
  personalized computational heart models,'' \emph{Europace}, vol.~25, no.~1,
  pp. 223--235, 2023.

\bibitem{xie2022advanced}
E.~Xie, E.~Sung, E.~Saad, N.~Trayanova, K.~C. Wu, and J.~Chrispin, ``Advanced
  imaging for risk stratification for ventricular arrhythmias and sudden
  cardiac death,'' \emph{Frontiers in Cardiovascular Medicine}, vol.~9, p.
  884767, 2022.

\bibitem{okada2020substrate}
D.~R. Okada, J.~Miller, J.~Chrispin, A.~Prakosa, N.~Trayanova, S.~Jones,
  M.~Maggioni, and K.~C. Wu, ``Substrate spatial complexity analysis for the
  prediction of ventricular arrhythmias in patients with ischemic
  cardiomyopathy,'' \emph{Circulation: Arrhythmia and Electrophysiology},
  vol.~13, no.~4, p. e007975, 2020.

\end{thebibliography}

\end{document}